\documentstyle[pra,aps,epsfig,twocolumn]{revtex}

\begin{document}
\twocolumn[\hsize\textwidth\columnwidth\hsize\csname@twocolumnfalse\endcsname
\title{Theoretical analysis of quantum dynamics in 1D lattices: 
Wannier-Stark description}
\author{Quentin Thommen, Jean Claude Garreau and V{\'e}ronique Zehnl{\'e}}
\address{Laboratoire de Physique des Lasers, Atomes et Mol{\'e}cules,
  UMR 8523, \\ and
Centre d'Etudes et de Recherches Laser et Applications,\\
Universit\'{e} des Sciences et Technologies de Lille, \\
F-59655 Villeneuve d'Ascq Cedex, France\\
}
\date{\today}
\maketitle

\begin{abstract}
This papers presents a formalism describing the dynamics of
a quantum particle in a one-dimensional tilted time-dependent lattice.
The description uses the Wannier-Stark states, 
which are localized in each site of the lattice and provides
a simple framework leading to
fully-analytical developments. Particular attention is devoted to 
the case of a time-dependent potential, which results in a rich variety of 
quantum coherent dynamics is found.\\
\vspace{0.5cm}
Pacs number(s): 03.65.-w, 03.75.-b, 32.80.Lg, 32.80.Pj
\end{abstract}
]

\section{Introduction}
\label{sec:intro} 
Quantum dynamics in a periodic lattice is one of the oldest 
problems of quantum mechanics, whose basis have been
settled by Bloch and Zener \cite{ref:Bloch,ref:Zener}, in the 30's. Aimed at 
the description of the electron motion in crystalline lattices,
this problem has largely been considered, for about half a century, as
an academic one, because dissipation effects forbid the observation
of most quantum effects in the motion of a crystalline electron.
Laser cooling of atoms has brought a revival of the interest on such
problems, as it produces atoms
whose de Broglie wavelength is comparable to the wavelength of the light
interacting with the atoms, and whose resulting kinetic energy is comparable
to the typical lightshift induced by the radiation. The latter
feature means that the cold atoms can be trapped in light potentials (or dipole
potentials). The former means that the atom dynamics in such a potential
is, in absence of dissipation, essentially quantum. Moreover, the
main source of dissipation is spontaneous emission, which can be arbitrarily
reduced (if one disposes of a powerful enough laser), whereas keeping a
constant light potential, just by an increase of the laser-atom detuning
\cite{ref:LesHouches}.

Light potentials are a consequence of the displacement of atomic levels
resulting from the interaction with light, corresponding to a process in
which a photon is absorbed transferring the atom to an (virtual) 
excited state from which the atoms de-excites back to the
ground-state by {\it stimulated} emission. Such a process induces an
energy shift
of the atomic levels that can be deduced from second order perturbation
theory and which is proportional to light intensity, to the square of the
coupling (that is, to $|{\bf d_{eg} \cdot \epsilon}|^2$, where ${\bf d_{eg}}$
is the dipole matrix element between the states $g$ and $e$, and $%
{\bf \epsilon}$ is the polarization vector of the light), and to the inverse
of the laser-atom detuning $\delta_L=\omega_L-\omega_{eg}$
($\omega_{eg}$ is the Bohr frequency between states $g$ and
$e$). Any spatial
gradient of this energy shift produces a (conservative) force, and thus a
potential. A simple example is that of a standing wave formed by two
counter-propagating parallel-polarized beams. Placed in such a standing wave,
an atom perceives a periodic one-dimensional potential whose strenght
varies sinusoidally in the space. Standing waves (with little variations)
form the model potential considered in the present work.

Light potentials generated by standing waves have been used
in many experimental studies of quantum dynamics. For example, Bloch
oscillations have been observed both with single atoms \cite{ref:BlochOsc} and
with a Bose-Einstein condensate (BEC) \cite{ref:BlochOscBEC} in an accelerated
standing wave. Wannier-Stark ladders \cite{ref:WSLadders} and collective
tunneling effects \cite{ref:CollTunnel} have also been studied with such
a system. Atoms placed in an intense,
phase-modulated, or pulsed, standing wave realize a paradigmatic system for
theoretical and experimental studies of quantum chaos, the so-called
Quantum Kicked Rotor
\cite{ref:QKRRaizen,ref:QKRChristiensen,ref:QKRLille,ref:QKROxford}.

In this paper, we consider the quantum dynamics of an atom (of mass $M$)
placed in a tilted sinusoidal potential whose phase (that is, the position of
its nodes) can be modulated in an arbitrary way, corresponding to
the Hamiltonian 
\begin{equation}
  H={\frac{p^{2}}{2M}}+v_{0} \cos \{  2k_{L}
      \left[x-x_{0}(t)\right] \} +f(t)x
\label{eq:H}
\end{equation}
where $x_{0}(t)$ is a  phase and $f(t)$ a
force, both being (eventually) time-dependent, and
$k_{L}=2\pi/\lambda_L$ is the wavenumber
of the standing wave. Different temporal
dependences of $x_{0}(t)$ can be considered. For instance, the
accelerated case, $x_{0}(t)=(1/2)at^{2}$, which has been studied in \cite
{ref:BlochOsc,ref:WSLadders,ref:CollTunnel},
is equivalent to an inertial force $F=Ma$ in the frame of the
potential (see appendix \ref{app:UnitTransf}).

A natural energy unit in such a context is the ``recoil energy'',
defined as the change in kinetic energy of the atom corresponding 
to the absorption a photon, given by 
\begin{equation}
E_{R}={\frac{\hbar ^{2}k_{L}^{2}}{2M}}  \label{eq:RecoilEnergy}
\end{equation}
to which one can associate a recoil frequency $\omega _{R}=E_{R}/\hbar $ and
a recoil momentum $p_{R}=\hbar k_{L}$, etc. It is also useful to re-scale the
variables: $X \equiv x/(\lambda_L/2)$, where $\lambda_L/2$ is the step of the
periodic lattice, and $\tau \equiv \omega _{R}t$. With these
definitions the above Hamiltonian takes the following form, which is
retained in the rest of the paper:
\begin{equation}
H={\frac{P^{2}}{2m^*}}+V_{0}\cos \{ 2\pi[X-X_{0}(\tau )]\}+F(\tau)X
\label{H_ren}
\end{equation}
where $V_{0} \equiv v_{0}/E_{r}$, $F \equiv f\lambda_L/ 2E_R$
(note that in this
system the momentum operator in the real space is
$P=-i(\partial/\partial X)$, $\hbar=1$, the reduced mass is
$m^*=\pi^2/2$ and $d=1$ is the step of the lattice).
For simplicity, in what follows, we shall write the rescaled
variables $x$, $p$ and $t$.

In the next section, we briefly review the time-independent Hamiltonian case
in Eq.~(\ref{H_ren}), introduce the
Wannier-Stark basis, and show that it leads to a very simple description
of the Bloch oscillation. In the following sections, we shall discuss the
more complicated dynamics that arises when a harmonic modulation of the lattice is
applied.

\section{The Bloch oscillation in the Wannier-Stark description}
\label{sec:WSBasis} 
The Bloch oscillation (BO) is a well known phenomenon 
discovered by Zener while studying the quantum properties of an electron in
a (perfect) crystal submitted to a constant electric field \cite{ref:Zener}.
The BO arises when a small spatial tilt is added to the lattice:
Quantum particles do not fall
along the slope of the potential, but perform a periodic, space-limited,
oscillation. BO is thus a strictly quantum behavior. We shall call
``lattice'' the untilted potential, and ``tilted lattice'' 
the sum of the lattice and the tilted potential.

The BO is usually described in the basis of the so-called {\em Bloch
states} \cite{ref:Bloch}, 
i.e., the eigenstates of the lattice.
In this paper, we use another framework corresponding to
the eigenstates of the {\em tilted} potential. While the lattice
is invariant under spatial translations by a multiple of
the spatial period $d$, the tilted lattice has a more complicate symmetry:
it is invariant under simultaneous spatial translation by a lattice
period $d$ {\em and}
energy translation by $\omega_B = Fd$ ($\omega_B$ is the ``Bloch
frequency''). One then expects the eigenenergies to form
``ladder'' structures separated by $\omega_B$, the so-called Wannier-Stark
ladders, introduced by Wannier in connection with
the problem of electrons in a crystal submitted to a homogeneous electric
field \cite{ref:Wannier}.
Each element of the ladder corresponds to eigenfunctions (the
Wannier-Stark states) centered at a given well, and thus separated 
by an integer multiple of $d$. The
form, and even the existence of these eigenstates has been the object of a
long controversy, that has been settled only recently \cite{ref:Nenciu}. In
the present paper, we consider a spatially limited lattice, extending over many
periods, and limited by an infinite-height box. This changes only
very slightly the ``bulk'' properties of the system, and the 
eigenenergies and eigenstates obtained numerically display (to a very good
approximation) the expected ladder structure described above.
In this framework, Wannier-Stark states (WSS) are
the eigenfunctions of the system.
The relation to the case of an infinite 
tilted lattice makes no problem if the corresponding states
(Wannier-Stark resonances) have long enough life-times compared
to the experimental times. The existence of these states has been evidenced in 1988 in a
semi-conductor superlattice \cite{ref:WSLaddersSL}, and 1996 with cold atoms in
an optical lattice \cite{ref:WSLadders}. Note that, being
spatially localized, Wannier-Stark states provide an ideal tool for
the description of the wave function of a cold atom (as produced by
a ``Sisyphus-boosted" MOT) placed in the
potential, whose de Broglie wavelength (around $\lambda_L/3$) is
of the order of the lattice period $\lambda_L/2$. 

Consider the properties of the time-independent Hamiltonian $H_{0}$ 
\begin{equation}
H_{0}={\frac{p^{2}}{2m^*}}+V_{0}\cos (2\pi x)+Fx 
\label{eq:H0}
\end{equation}
where $F$ is a constant force. The eigenfunctions of $H_{0}$ are
Wannier-Stark states forming an energy ladder whose separation is
$\omega_B=Fd$, $E_{nm}=E_m+nFd$.

\vspace{0.5cm}
\begin{figure}[tbp]
  \begin{center}
    \psfig{figure=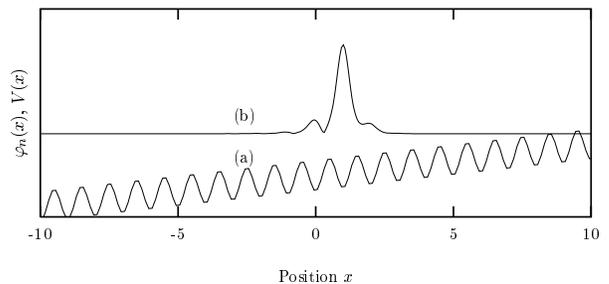,width=8cm,clip=}
\caption{(a) Periodic potential with a tilt. (b) The WSS
  state $\varphi_n$ is localized in the well of index $n=1$
  and has appreciable overlap with neighbor lattice sites $n=0,2$.
  This eigenfunction is obtained for $V_0=2.5$ and $F=0.5$. 
The ``numerical'' box includes 64 lattice sites.}
\label{fig:TiltedPotential}
\end{center}
\end{figure}
\vspace{0.5cm}

The BO can be advantageously described by using
the Wannier-Stark state (WSS) localized inside a given 
individual lattice well, corresponding to the lowest energy of
the states associated to this well (see Fig.~\ref{fig:TiltedPotential}).
Note that considering only the
ground state of each well is equivalent to the restriction to
the first Bloch band in the description based on Bloch states.
We also choose strong enough $F$ and $V_0$ to produce well-localized WSS
\cite{ref:Fukuyama}. The WSS associated to the lattice well labeled
$n$ is noted $\varphi_{n}(x)$ (supposed real) and the corresponding
eigenenergy is $E_{n}$ (we drop the index $m$). The symmetries of the
potential discussed above then imply:
\begin{equation}
 \varphi _{n+p}(x) = \varphi _{n}(x-pd)
\label{eq:WSSTranslation}
\end{equation}
and
\begin{equation}
E_{n+p}=E_{n}+ pFd \text{ .}
\label{eq:EnergyTranslation}
\end{equation}

Let us describe the atomic wave function by a superposition of WSS 
\begin{equation}
\Psi (x,t)=\sum_n c_{n}(t)\varphi _{n}(x)
\label{eq:Psi_WS}
\end{equation}
with $c_n(t) = c_{n} e^{-iE_{n}t}$; $c_n$ being the amplitude at $t=0$.
The dynamical quantities of the atom can easily calculated. For
instance, the mean value of the atomic position operator is: 
\begin{equation}
\langle x\rangle =\sum_n X_{n,n}\left| c_{n}\right| ^{2}+
\sum_{n\ < m} \left( X_{n,m}c_{n}^{\ast }c_{m}e^{i(E_{n}-E_{m})t}+c.c \right)
\label{eq:xmean}
\end{equation}
where $X_{n,m} \equiv \left\langle \varphi _{n}\left| x\right| \varphi
_{m}\right\rangle$. As long as the WSS $\varphi_n$ is well localized in
the respective well, we can keep only
nearest-neighbors contributions ($n=m\pm 1)$ and derive a simplified
expression: 
\begin{equation}
  \langle x\rangle =\overline{x}+ X_{n,n+1} \left( \sum_{n}c_{n}^{\ast
      }c_{n+1}
    e^{-i\omega_{B}t}+c.c \right)
\end{equation}
where $\overline{x} = \sum X_{n,n} \left| c_{n}\right|^{2} $ is the mean
position and $X_{n,n+1}=X_{0,1}=X_{0,-1}=X_{n,n-1}$ is independent of
$n$ \cite{note:Identities}. This
result evidences a counter-intuitive property of the quantum
motion in a tilted lattice: instead of falling along the slope,
the atom performs an oscillation
with frequency $\omega_B$ (the Bloch frequency). The amplitude of this 
Bloch oscillation is proportional to $X_{0,1}$ and grows with the overlap
between neighbors WSS, i.e for a small slope $F$
and small lattice depth $V_{0}$. The physical origin of the BO
appears here clearly as an interference effect between neighbor sites, as
$c_n^* c_{n+1}$ is the coherence between the sites $n$ and 
$n+1$. Note the lack of oscillations if
the atom is localized in only one well. 
Eq.~(\ref{eq:xmean}) also predicts the occurrence 
of harmonics with frequency $p\omega_B$ ($p$ integer)  
but with smaller amplitudes since they involve the coupling 
strength $X_{n,n+p}$ \cite{note:amplitude}.

The description of the BO given here is
very different of the usual ``solid-state'' approach. There, the Bloch
states (eigenstates of the untilted lattice)
are taken as the reference basis. The oscillation is described in a
semi-classical frame as the periodic evolution of the atom's
quasi-momentum $Ft$ (in the first Brillouin zone) with period $T=2\pi/(Fd)$ 
\cite{ref:Mermin,ref:Holthaus}. 
Although intuitive, this approach does not make clear the
role of quantum interference produced by the periodic lattice
structure as the basic
mechanism underlying the Bloch oscillation, which is evidenced in our approach.

\section{The modulated potential in the Wannier-Stark description:
  Resonant dynamics}
\label{sec:TimeDepWSS}

With the existence of the natural frequency $\omega_B$ of the system
in mind, one is tempted
to investigate the quantum dynamics in presence of a harmonic
external forcing. The WSS approach proves to be very efficient,
since it allows a fully-analytical description.
After some general considerations, we study in this section the case
of resonant forcing, and show that it leads to a very rich and
interesting dynamics. The general (non-resonant) case will be treated in the
next section.

Consider the Hamiltonian of Eq.~(\ref{H_ren}) with a constant force 
$F$ and a lattice phase modulation
\begin{equation}
x_{0}(t )=a\sin (\omega t) \text{.}
\label{eq:modulation}
\end{equation}
The developments are simpler if we use a unitary transformation that
transforms the modulation in a time-dependent force. Physically, 
this is equivalent to move to an accelerated
reference frame in which the lattice is at rest \cite{ref:BlochOsc,ref:Ben},
adding thus an inertial force (Appendix \ref{app:UnitTransf}). In this 
frame, the new Hamiltonian is given by Eq.~(\ref{eq:H0}) plus a
time-dependent force $F'(t)$:
\begin{equation}
F^\prime(t )=m^*\frac{d^{2}x_{0}(t )}{dt^{2}}=
-m^*a\omega^{2}\sin (\omega t ) = -F_0 \sin(\omega t)
\label{eq:Finertial}
\end{equation}
where $F_0 \equiv m^*a\omega^2$ is the amplitude of the inertial force.
The harmonic time-dependence in Eq.~(\ref{eq:Finertial}) is the analog of an
AC electric field for electrons in a (perfect) crystal. The dynamics
in such a system
can be described in a quite simple fashion by writing the state of the atom
as a superposition of WSS, Eq.~(\ref{eq:Psi_WS}). The coefficients
$c_n(t)$ can be obtained by reporting Eq.~(\ref{eq:Psi_WS})
into the Schr{\"o}dinger equation 
\begin{equation}
[ H_0-F_0 x \sin (\omega t) ] \Psi(x,t) = 
i { \partial \Psi(x,t) \over \partial t}
\label{eq:ModulatedH}
\end{equation}
where $H_0$ is given by Eq.~(\ref{eq:H0}), with eigenstates 
$\varphi_n(x)$. This produces the following set of
coupled differential equations for the $c_n(t)$:
\begin{equation}
\dot{c}_{n}(t)=-iE_{n}c_{n}(t)+i F_0 \sin(\omega t)
 \sum_{m}X_{n,m}c_{m}(t)
\end{equation}
where $\dot{c}_n \equiv dc_n/dt$. 
Neglecting temporarily the coupling 
between different WSS, (i.e putting $X_{n,m}=X_{n,n} \delta_{m,n}$),
the amplitudes are obtained simply as $c_n=\exp [i\phi_n(t)]$, with the
time-dependent phase 
\begin{equation}
\phi_n(t)=-E_n t-{F_0 X_{n,n} \over \omega} \cos(\omega t)
\label{eq:phi_n}
\end{equation}
where
$X_{n,n}=X_{0,0}+nd$, depends on the site index $n$ \cite{note:Identities}.
We now write $c_n(t)\equiv d_n(t) e^{i\phi_n(t)}$. The amplitudes
$d_n$ obey the following system of differential equations:
\begin{equation}
\dot {d}_{n}=i F_0 \sum_{m \neq n}X_{n,m}d_{m}(t)
\exp \{ i \left[\phi_{m}(t)-\phi_{n}(t) \right]\} \sin(\omega t)
\text{ .}
\label{eq:dn0}
\end{equation}
After Eq.~(\ref{eq:phi_n}), the phase difference $\phi_m(t)-\phi_n(t)$ is:
\begin{equation}
\phi_{m}(t)-\phi_{n}(t)=(n-m) \left[\omega_Bt +{ F_0 d \over \omega }
\cos(\omega t) \right]
\end{equation}
where we used Eq. (\ref{eq:EnergyTranslation}).
Eq.~(\ref{eq:dn0}) can be recast as
\begin{mathletters}
\begin{equation}
\dot {d}_{n} =
i F_0 \sum_{p \neq 0} X_p d_{n+p} \left[ e^{-ip\omega_Bt}
e^{-i(pF_0d/\omega) \cos(\omega t)} \right] \sin(\omega t)
\label{eq:dn1}
\end{equation}
\begin{eqnarray}
= { F_0 \over 2 } \sum_{p \neq 0}X_p d_{n+p} 
\sum_{l} (-i)^l J_l \left(p{ F_0 d\over \omega}\right) \nonumber \\
\{ e^{ i \left[(l+1)\omega  -p\omega_B  \right] t }- 
e^{ i \left[(l-1)\omega  -p\omega_B  \right] t }\}
\label{eq:dn2}
\end{eqnarray}
\label{eq:dn}
\end{mathletters}
where, $X_p\equiv X_{n,n+p}$ \cite{note:Identities}. $J_n(x)$ is the Bessel
function of the first kind, and we have used the well-known property
of Bessel functions:
\begin{equation}
  e^{-iz \cos(\omega t)} = \sum_{l=-\infty}^{+\infty}
  J_l(z) (-i)^l e^{il\omega t} \text{ .}
\label{eq:BesselGenerator}
\end{equation}
Note that the sum over lattice sites (i.e over $p$) extends
only over a few neighbors sites, since the coupling coefficients $X_p$ rapidly 
shrink to zero \cite{note:amplitude}. If the
modulation is smooth enough to avoid projections on other 
states of the system,  the sum over 
the harmonics of the modulation (i.e. $l$) is also limited to a few terms
close to $l=0$ (typically, $l_{max} \sim p F_0 d/ \omega =
pm^*a\omega d \sim O(1)$).
Therefore, only a finite number of terms are to be retained in the above 
expression. On the other hand, the evolution of $d_n$ described by
Eqs. (\ref{eq:dn2}) is a sum of oscillations with frequencies
$\left[(l\pm 1)\omega-p\omega_B \right]$. In the following we keep
only the so-called {\em secular} terms, that is, 
terms that oscillate slowly or do not oscillate at all. The resulting
``close to resonance" dynamics is observed when $(l\pm 1)\omega 
\approx p\omega_B $, i.e when the forcing frequency $\omega$ is
commensurable (or almost) with the system's natural frequency $\omega_B$. 

Let us consider the simpler resonant case, $\omega = \omega _{B}$.
Due to the relative strength of the factors $X_{p}$ we can keep, 
to a good accuracy, only the contribution of the
next-neighbor site ($p=1$), which leads to the 
following expression: 
\begin{equation}
 \dot {d}_{n}(t)=\Omega_1 \left[ d_{n+1}-d_{n-1} \right]
\label{eq:dnsimple}
\end{equation}
where
\begin{eqnarray}
\Omega_1 = { F_0 X_1 \over 2 } \left[ 
  J_0\left( { F_0 d \over \omega_B } \right)
  + J_2\left({ F_0 d \over \omega_B }\right) \right] \nonumber \\
= { \omega_B X_1 \over d } J_1 \left({F_0 d \over \omega_B} \right){ .}
\label{eq:Omega1}
\end{eqnarray}
This equation is similar to a ``dipole coupling" between sites
$n$ and $n \pm 1$ where $\Omega_1$ plays the role of a Rabi frequency.
Note that, contrary to intuition, the coupling towards the left
or towards the right neighbor is the same.

The meaning of Eq.(\ref{eq:dnsimple}) can be better appreciated
by searching for the plane-wave solutions of the form: 
\begin{equation}
d_{n}(t)=e^{i\left( k_{0}d n+\omega t\right) } \text{ .}
\label{eq:WSWave}
\end{equation}
The (dimensionless) wavenumber $k_0$ takes into account the 
phase difference between neighbor
sites. Substitution into Eq.~(\ref{eq:dnsimple}) 
leads to the dispersion relation 
\begin{equation}
\omega =2 \Omega_1 \sin(k_{0}d)  
\label{eq:dispers}
\end{equation}
and to the group velocity $v_{g}\equiv d\omega /dk_{0}:$ 
\begin{equation}
v_{g}=2\Omega_1 d\cos(k_{0}d) \text{ .}  
\label{eq:vg}
\end{equation}
This result shows that the dynamics is different
depending on the wave number $k_{0}$. For instance, if $k_{0}d=\pm \pi /2$
(phase-quadrature from site to site), $v_{g}=0$ and there is no global motion. 
If $k_{0}d=\pi$, $v_{g}=-2\Omega_1 d$ and the global motion is a fall along
the slope of the potential with the maximum speed
$2\Omega_1 d$. More interesting is the case $k_{0}=0$, where
$v_{g}=2\Omega_1 d$: the atom then climbs up the slope of the potential with
a constant maximum speed: there is, in this case, {\em coherent transfer
of energy} from the modulation to atom, thanks to 
the particular phase relations between neighbor sites.  Note also that, 
contrary to the motion of a classical particle, the speed 
$\left| v_{g}\right|$ is independent of the sense of displacement:
the wavepacket climbs the slope up or down at the same speed.

More detailed information on the wavepacket motion can be grabbed by
writing the amplitude $d_n$ in the more general form: 
\begin{equation}
d_{n}(t)=f_{n}(t)e^{i\left( k_{0}dn+\omega t \right) }
\end{equation}
where $f_{n}$ are complex amplitudes describing the envelope of the 
atomic wavepacket, assumed to vary slowly in time as 
compared to the frequency  $\omega_B$, and in space as compared
to the lattice period $d$. Reporting the
above expression into Eq.~(\ref{eq:dnsimple}), we get: 
\begin{eqnarray}
\dot{f}_{n}+i\omega f_{n}=\Omega_1 \left[ \cos (k_{0}d)\left(
    f_{n+1}-f_{n-1}\right) \right. \nonumber \\
  \left. +i\sin (k_{0}d)\left(f_{n+1}+f_{n-1}\right) \right] 
\text{ .}
\end{eqnarray}
Using the dispersion relation Eq.~(\ref{eq:dispers}) and
keeping only slowly varying contributions, we obtain: 
\begin{eqnarray}
  \dot {f}_{n}=\Omega_1 \left[ \cos (k_{0}d)\left(
  f_{n+1}-f_{n-1}\right) \right. \nonumber \\
\left.
  +i\sin (k_{0}d)\left( f_{n+1}+f_{n-1}-2f_{n}\right) \right] 
\text{ .}
\label{eq:slowfn}
\end{eqnarray}
Since $f_{n}$ vary slowly in space, we can take the continuous limit
(with respect to the variable $x=nd$) and deduce the following 
equation for the wavepacket envelope: 
\begin{equation}
  \dot{f}(x,t) =\Omega_1 \left( 2d\cos (k_{0}d) {\frac{\partial
        }{\partial x}}
    +id^2\sin(k_{0}d) {\frac{ \partial^2 }{\partial x^2 }} \right) f(x,t)
\text{ .}
\label{eq:PDE}
\end{equation}
This equation is an interesting piece of information. If 
$k_{0}d=\pm \pi/2$, one has a diffraction equation of the form $\dot{f}
=\pm i\Omega_1 d^2 \partial_{x}^{2}f$, which describes the spreading of
the atomic wavepacket (i.e. diffraction) with no global
displacement. The case $k_{0}d=0,\pi$ gives the wave equation $%
\dot{f}=v_g \partial_x f$ ($v_g=2d\Omega_1$), describing a
wavepacket traveling with constant velocity $v_{g}$ and no deformation: the
wavepacket presents an ascending or descending coherent motion. Mixed
behaviors, i.e. spreading at a diffusion rate $\Omega_1 d^2 \sin (k_{0}d)$
and uniform displacement with group velocity $v_g=2\Omega_1 d\cos (k_0d)$,
are found for other values of $k_{0}$. The general
solution can be easily obtained by performing a spatial Fourier transform
of Eq.~(\ref{eq:PDE}).

\vspace{0.5cm}
\begin{figure}[tbp]
  \begin{center}
    \psfig{figure=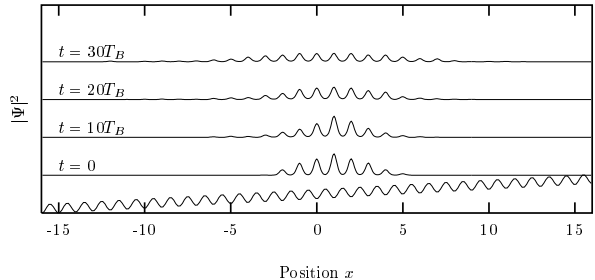,width=8cm,clip=}
  \caption{The atomic
    wavepacket is obtained from a full integration of the
    Schr{\"o}dinger equation corresponding to the modulated
    tilted lattice. The wavepacket at $t=0$ has a site-to-site
    phase of $\pi/2$. The wavepacket spreads in time with
    no displacement, in an agreement with the theoretical prediction. 
    Parameters are $V_0=2.5$, $F=0.5$,  $\omega=\omega_B=0.5$
    and $a=0.2$}
  \label{fig:ResDynQuadrature}
  \end{center}
\end{figure}
\vspace{0.5cm}
\begin{figure}[tbp]
  \begin{center}
    \psfig{figure=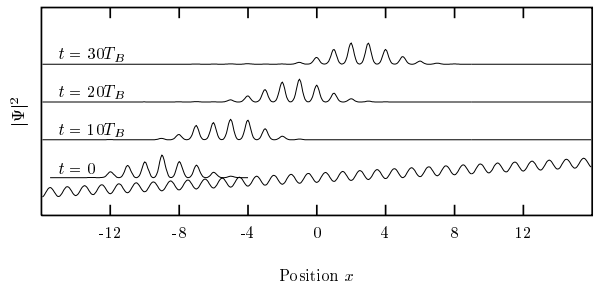,width=8cm,clip=}
  \caption{Same as Fig.~\protect\ref{fig:ResDynQuadrature}, except
    that the site-to-site phase is $k_0=0$. The wavepacket has a
    shape-preserving motion upwards the slope of the potential. The
    observed group velocity is 0.030, in good agreement with the
    prediction of Eq.~(\protect\ref{eq:vg}), $v_g=0.032$  ($X_{0,1}=0.13$).}  
  \label{fig:ResDynInPhase}
  \end{center}
\end{figure}
\vspace{0.5cm}

The above result shows that, depending on the initial wavenumber $k_{0}$,
i.e. on the way the initial wave packet is prepared, the atom behaves in
qualitatively different ways. Figs. \ref{fig:ResDynQuadrature} and
\ref{fig:ResDynInPhase} are obtained by a direct
integration of the Schr\"{o}dinger equation, with the Hamiltonian given by
Eq.~(\ref{eq:ModulatedH}), for different evolution times. 
In the first case the initial wavepacket is prepared
with site-to-site phase  $k_0d=\pi/2$ and diffraction is clearly visible, 
as predicted. In Fig.~\ref{fig:ResDynInPhase}, the wavepacket, prepared
with $k_0=0$ has a uniform displacement while preserving its shape.
Its group velocity obtained from the numerical simulations is $v_g=0.030$,
in very good agreement with the theoretical 
value from Eq.~(\ref{eq:vg}) which is $v_g=0.032$.
\vspace{0.5cm}
\begin{figure}[tbp]
    \begin{center}
    \psfig{figure=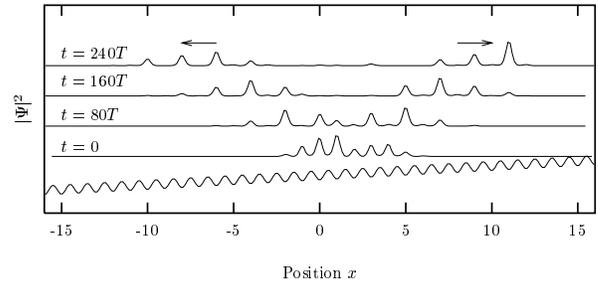,width=8cm,clip=}
\caption{Evolution of a wave packet for $\omega=2\omega_B$. The initial
  wavepacket is a superposition of $\varphi_n$ ($n$ odd) which are
  in phase, and of $\varphi_n$ ($n$ even) with phase difference $\pi$.
  The initial wavepacket separates in two packets with opposite velocities.}
\label{fig:HarmonicDyn}
\end{center}
\end{figure}
\vspace{0.5cm}
Other resonant behaviors are observed if $\omega=q\omega_B$ 
($q$ integer). From the general expression of Eq.~(\ref{eq:dn}), 
one finds for instance a next-to-neighbor ($n \rightarrow n\pm 2$)
resonant interaction if $\omega=2\omega_B$, leading to: 
\begin{equation}
\dot{d}_{n}(t)=\Omega_2  \left[ d_{n+2}-d_{n-2}\right]
\label{eq:DynNextNeig2}
\end{equation} 
with $\Omega_2 = (\omega_B/d) X_2 J_1(F_0 d/\omega_B)$. 
Note that $X_{2} \ll X_1$ \cite{note:amplitude}. 
We illustrate in Fig.~\ref{fig:HarmonicDyn} the temporal evolution
of a wavepacket, obtained by numerical integration of the Schr\"{o}dinger
equation. The initial state is prepared as a superposition
of two packets: one packet is constructed with in-phase amplitudes  
(that is, $c_n$ and $c_{n+2}$ have the same phase for $n$ odd),
and the other one is constructed with amplitudes in phase-opposition
(that is, $c_n$ and $c_{n+2}$ have opposite sign for $n$ even).
The first packet
moves with velocity $v_g=4\Omega_2 d$, and the second with $v_g=-4\Omega_2 d$. 
The figure displays an original behavior showing each of these
two initially inter-penetrated packets moving independently
in opposite directions, creating a highly delocalized state.

\section{The modulated potential: general case}
\label{sec:GeneralCase}
In this section we generalize the results of the preceding section
to the case of a non-resonant modulation. We follow
essentially the same steps as in Sec.~\ref{sec:TimeDepWSS}, and
we shall skip algebraic details of the calculations.
Coming back to Eq.~(\ref{eq:dn1}) and looking for a solution of the form: 
\begin{equation}
d_n(t)=e^{i[(k_0 d n +\phi(t)]} 
\label{eq:WaveHR}
\end{equation}
one gets the instantaneous frequency:
\begin{eqnarray}
  \dot{\phi}
  &=& 2 F_0 \sum_{p > 0}X_{p}
  \cos \{p[k_0 d -\theta(t)]\}   \sin(\omega t)
 \end{eqnarray}
where $\theta(t) = \omega_Bt +(F_0 d/\omega) \cos(\omega t)$, and
we used $X_{p}=X_{-p}$. The group velocity
$v_g=d\dot{\phi}/dk_0$ is thus
\begin{equation}
  v_g= 2F_0 d  \sum_{p > 0}pX_{p} \sin \{p[\theta(t)-k_0 d]\}
  \sin(\omega t)
\label{eq:vegeneral}
\end{equation}
As in the preceding section, more detailed behavior is
obtained by putting
\begin{equation}
d_n(t)=f_n(t)e^{i[k_0dn+\phi(t)]}
\end{equation}
where $f_n$ are slowly-varying amplitudes.
The generalization of Eq.~(\ref{eq:slowfn}) is then:
\begin{equation}
\dot{f}_{n}=
i F_0 \sum_{p \neq 0} X_{p} [f_{n+p}-f_n] e^{ip(k_0d-\theta)}
\sin(\omega t)
\end{equation}
or,
\begin{eqnarray}
\dot{f}_{n} = F_0 \sum_{p > 0}X_{p} \sin(\omega t) \nonumber \\
\{
    - \left[f_{n+p}-f_{n-p} \right] \sin[p(k_0d-\theta)] \nonumber  \\ 
  +i\left[ (f_{n+p}+f_{n-p}-2f_n) \right] \cos[p(k_0d-\theta)]   \} \text{ .}
\end{eqnarray}
Taking the continuous limit of the above expression then
produces an equation describing both the propagation and the
diffraction of the wavepacket:
\begin{eqnarray}
\dot{f}(x,t) =\left(v_g(t) {\frac{\partial }{\partial x}} 
  +iD(t) {\frac{ \partial^2 }{\partial x^2 }} \right) f(x,t) \nonumber\\
+2i F_0  \sum_{p \ge 2}  X_p \cos[p(\theta-k_0d)]\sin(\omega t) f(x,t)
\label{eq:PDEHR}
\end{eqnarray}
where 
\begin{equation}
D(t)= F_0 d^2 \sum_{p>0} p^2 X_p \cos[p(\theta-k_0d)]
\sin(\omega t) 
\end{equation}
and the group velocity $v_g$ is given by Eq.~(\ref{eq:vegeneral}). 
Note that the last term in Eq.(\ref{eq:PDEHR}) is a phase term which
is $O(X_2)<<1$ and does not
contribute to the probability density $|f(x,t)|^2$. For the sake of
lightness, it is not considered in the following. 

The Fourier transform of Eq.~(\ref{eq:PDEHR}) with respect to $x$ 
produces an  algebraic equation for the Fourier transform
$\tilde{f}(k,t)$ of $f(x,t)$, whose solution is:
\begin{equation}
  \tilde{f}(k,t) = e^{ik x^\prime(t)} e^{ik^2 \Delta(t)} \tilde{f}(k,0)
\end{equation}
and thus:                 
\begin {equation}
  f(x,t)= { 1 \over \sqrt{2\pi} } \int e^{ik[x+x^\prime(t)]} e^{ik^2 \Delta(t)}
   \tilde{f}(k,0) 
\end{equation}
with
\begin{equation}
  x^\prime(t)= \int_0^t v_g(\tau) d\tau
\end{equation}
 and
\begin{equation}
  \Delta(t) = \int_0^t D(\tau) d\tau \text{ .}
\end{equation}
This expression describes a coherent motion of the wavepacket formed of
an oscillatory motion with the time-dependent group velocity
Eq.~(\ref{eq:vegeneral}), and a diffusive motion  with a time-dependent
diffusion coefficient $D(t)$. For example, if one builds an initial
gaussian packet of width $a_0$, $f(x,0)= \exp(-x^2/a_0^2)$
one finds, after some straightforward calculations:
\begin{equation}
  |f(x,t)|^2 =\frac{a_0}{a(t)}
   \exp\left(-{\frac{2x(t)^2}{a(t)^2}}\right)
   \text{ ,}
\end{equation}
where
\begin{equation}
  a(t)=a_0 \left[1+16{ \Delta(t)^2 \over a_0^4}
  \right]^{1/2} \text{ .}
\end{equation}
\vspace{0.5cm}
\begin{figure}[tbp]
  \begin{center}
    \psfig{figure=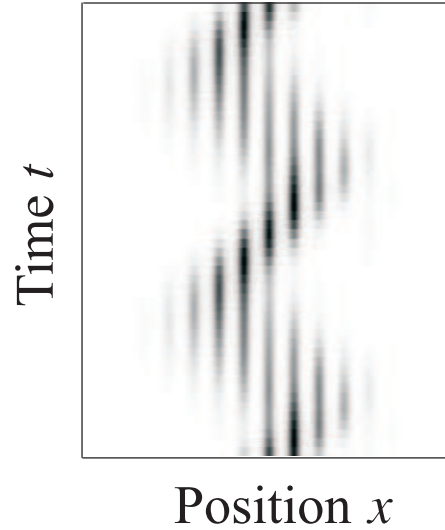,width=6cm,clip=}
  \caption{Spatio-temporal behavior of
    $|\Psi(x,t)|^2$ obtained numerically by integration 
    of the Schr\"{o}dinger equation (gray level convention: the maximum values
    of $|\Psi(x,t)|^2$ are depicted in black), from $t=0$ to
    $t=4\pi/\delta$ (that is, 2 periods of the beat frequency
    $\delta=\omega-\omega_B=0.02$). Other parameters are the same as in
    Fig.~\protect \ref{fig:ResDynQuadrature}. }
  \label{fig:EvolutionGeneral}
  \end{center}
\end{figure}
\vspace{0.5cm}
The physical meaning of our developments can be evidenced by
considering the case where $\omega$ differs from $\omega_B$ by
a small detuning $\delta=\omega -\omega_B$, $|\delta| \ll \omega_B$,
and keep only leading-order terms of order $O(\delta^{-1})$. 
The wavepacket then undergoes a harmonic oscillation at the beat frequency 
$\delta$ with a group
velocity given by Eq.~(\ref{eq:vegeneral})
\begin {equation}
   v_g(t)=2\Omega_1 d\cos(k_0d + \delta t)
\end{equation} 
corresponding to a periodic mean position displacement
\begin{equation}
  \langle x(t)\rangle=x(0)-\frac{2\Omega_1 d}{\delta}
  [\sin(k_0d + \delta t)-\sin(k_0d)]
\end{equation}
The width of the wavepacket oscillates in a breathing mode which
is governed by:
\begin{equation}
  \Delta(t) = \frac{\Omega_1 d^2}{\delta} [\cos(k_0d +\delta
  t)-\cos(k_0d)] \text{ .}
\end{equation}
The results of Sec. \ref{sec:TimeDepWSS} for a resonant excitation
are naturally recovered in the limit $\delta\rightarrow 0$.
\vspace{0.5cm}
\begin{figure}[tbp]
  \begin{center}
    \psfig{figure=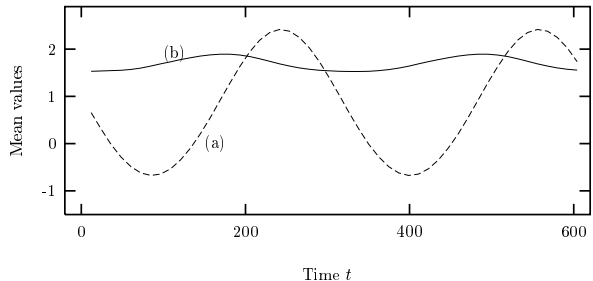,width=8cm,clip=}
    \caption{Mean values (a) $\langle x(t) \rangle$ and (b)
      $\sqrt{\langle x^2(t) \rangle - \langle x(t) \rangle^2}$ obtained
      from the wavepacket dynamics depicted in Fig.~\protect
      \ref{fig:EvolutionGeneral}.}
    \label{fig:MeanValues}
  \end{center}
\end{figure}
\vspace{0.5cm}
We have performed the integration of the Schr\"{o}dinger equation
for a detuning $\delta=0.02$. Fig.~\ref{fig:EvolutionGeneral} shows
the spatio-temporal dynamics 
of the wavepacket and clearly evidences the periodic oscillations
and the breathing at frequency $\delta$ predicted above. Figure 
\ref{fig:MeanValues} shows
the mean position and width of the wavepacket 
as a function of time. 
The comparison with the theory is very good. For instance, the
amplitude of $\langle x(t)\rangle$ is found numerically as 1.55
compared to the theoretical value 1.73.

\section{conclusion}
We have studied, in a fully-analytical way, the dynamics of a wavepacket in a
static and time-modulated tilted potential in the framework of the
Wannier-Stark states. This
basis is well suited to the description of
the state of a cold atom (as produced by a Sisyphus-boosted MOT).
Moreover, it provides a 
simple description the atomic dynamics, which is proved to be very rich: 
A variety of coherent motions are obtained depending
on the preparation of the initial wavepacket and its site-to-site
quantum coherence. We can note that the description
introduced here is, in its principle, independent of the details of
the lattice, provided it presents localized states in the lattice sites. 
It is therefore generalizable to other kinds of lattices.
 
Let us finally, mention that the present work distinguishes from
the more usual ``solid-state''
approach which is based on Bloch functions. 
We postpone for a forthcoming work the detailed comparison
between the Wannier-Stark and Bloch approaches. 

\section{acknowledgments}
Laboratoire de Physique des Lasers, Atomes et Mol{\'e}cules (PhLAM)
is UMR 8523 du CNRS et de l'Universit{\'e} des Sciences
et Technologies de Lille. Centre d'Etudes et Recherches
Lasers et Applications (CERLA) is supported by Minist\`{e}re de la
Recherche, R\'{e}gion Nord-Pas de Calais and Fonds Europ\'{e}en de
D\'{e}veloppement Economique des R\'{e}gions (FEDER).

\appendix
\section{Unitary transformation}
\label{app:UnitTransf}
Eq.~(\ref{eq:Finertial}) is  obtained if we perform a unitary transformation 
\begin{equation}
U(t)=e^{iX_{0}(t)P}e^{-i\beta (t)X}e^{i\gamma (t)}
\end{equation}
where we have included translation operators in space and momentum with $%
\beta =2m^*\dot{X}_{0}$(i.e momentum of a particle of mass $%
m^*$). In this framework, following \cite{ref:Ben}, we obtain (with $%
UXU^{+}=X+X_{0}$ and $UPU^{+}=P+\beta $) 
\begin{eqnarray}
H^{\prime } = UHU^{+}+i\left( d_{t}U\right) U^{+} \nonumber \\ 
= \frac{(P+\beta )^{2}}{2m^*}+V_{0}\cos (2\pi X) 
+F(X+X_{0}(t))- \dot{X}_{0}(t)P+ \nonumber \\
\dot{\beta }(t)\left( X+X_{0}\right) -
\dot{\gamma }  \nonumber
\end{eqnarray}
with $\dot{\gamma}=FX_{0}+m^* \stackrel{..}{X}_0 X_0+(m^*/2)
\dot{X}_{0}^{2}:$ 
\begin{equation}
  H^{\prime }=\frac{P^{2}}{2m^*}+V_{0}\cos (2\pi X)+
  (F+m^*\stackrel{..}{X_{0}}(t))X 
\end{equation}
Therefore, in the frame of the periodic potential, the Hamiltonian contains
an inertial force proportional to $\stackrel{..}{X_{0}}(t).$


\begin{references}

\bibitem{ref:Bloch} F. Bloch, Z. Phys {\bf 52}, 555 (1928).

\bibitem{ref:Zener}  C. Zener, Proc. R. Soc. London A {\bf 145}, 523
  (1934).

\bibitem{ref:LesHouches}  See for example {\it Fundamental Systems in 
Quantum Optics}, {\'E}cole d'{\'e}t{\'e} des Houches, Session  LIII, 1990,
edited by J. Dalibard, J. M. Raimond, and J.  Zinn-Justin (North-Holland,
Amsterdam, 1992).

\bibitem{ref:BlochOsc}  M. Ben Dahan, E. Peik, J. Reichel, Y. Castin,  and
C. Salomon, Phys. Rev. Lett. {\bf 76}, 4508 (1996).

\bibitem{ref:BlochOscBEC}  O. Morsch, J. H. M{\"u}ller, M. Cristiani,  D.
Ciampini, and E. Arimondo, Phys. Rev. Lett. {\bf 87}, 140402 (2001).

\bibitem{ref:WSLadders}  S. R. Wilkinson, C. F. Bharucha, K. W.  Madison, Q.
Niu, and M. G. Raizen, Phys. Rev. Lett. {\bf 76}, 4512 (1996).

\bibitem{ref:CollTunnel}  B. P. Anderson and M. Kasevich, Science {\bf \ 282}%
, 1686 (1998).

\bibitem{ref:QKRRaizen}  F. L. Moore, J. C. Robinson, C. F. Bharucha,  P. E.
Williams, and M. G. Raizen, Phys. Rev. Lett. {\bf 73}, 2974  (1994); B. G.
Klappauf, W. H. Oskay, D. A. Steck, and  M. G. Raizen, Phys. Rev. Lett. {\bf %
81}, 1203 (1998).

\bibitem{ref:QKRChristiensen}  H. Amman, R. Gray, I. Shvarchuck, and N. 
Christiensen, Phys. Rev. Lett. {\bf 80}, 4111 (1998).

\bibitem{ref:QKRLille}  J. Ringot, P. Szriftgiser, J. C. Garreau, and  D.
Delande, Phys. Rev. Lett. {\bf 85}, 2741 (2000).

\bibitem{ref:QKROxford}  M. B. D'Arcy, R. M. Godum, M. K. Oberthaler,  M. K.
Cassettari, G. S. Summy, Phys. Rev. Lett. {\bf 87}, 074102 (2001).

\bibitem{ref:Wannier} G. Wannier, Rev. Mod. Phys. {\bf 62}, 645 (1962).

\bibitem{ref:Nenciu}  For a complete review of the subject, see G.  Nenciu,
Rev. Mod. Phys. {\bf 63}, 91 (1991).

\bibitem{ref:WSLaddersSL}  J. Bleuse, G.Bastard and P. Voisin, Phys. Rev.
  Lett {\bf 60}, 220 (1988).

\bibitem{ref:Fukuyama} H. Fukuyama, R. A. Bari, and H. C. Fogdeby,
  Phys. Rev. B {\bf 8}, 5579 (1973).

\bibitem{note:Identities}  
By virtue of the translational
properties of the WSS, Eqs.~\protect(\ref{eq:WSSTranslation}) and
(\protect\ref{eq:EnergyTranslation}), $X_{n,n+p}$ ($p \neq 0$) 
does not depend on $n$:
$X_{n,n+p} = \int \varphi_n^*(x) x \varphi_{n+p}(x) dx =
\int \varphi_0^*(x-nd) x \varphi_{p}(x-nd) dx =
\int \varphi_0^*(x^\prime) (x^\prime+nd) \varphi_{p}(x^\prime)
dx^\prime = X_{0,p}$, where we used the orthogonality of the WSS.
Note however that $X_{n,n}=X_{0,0} + nd$ does depend on $n$.

\bibitem{note:amplitude} While the nearest neighbors interaction is
  proportional to $X_{01}=0.13$ the next to neighbor interaction is
  $X_{02}=7.8$ $10^{-3}$ i.e the amplitude of the oscillation at
  $2\omega_B$ is roughly 20 times smaller than that at $\omega_B$ (the
  numerical values were obtained for $F=0.5$ and $V_0=2.5$).

\bibitem{ref:Mermin}  N. W. Ashcroft N.W and N. D. Mermin,
  {\it Solid State Physics}, Holt Rinehart and Winston, 1976.

\bibitem{ref:Holthaus}  M. Holthaus, J. Opt. B: Quantum Semiclass.
  Opt. {\bf 2}, 589 (2000).

\bibitem{ref:Ben} M. Ben Dahan, {\it Th{\`e}se de doctorat} (unpublished)
  Paris, 1997.

\end{references}
\end{document}